\documentclass{emulateapj}
\usepackage{aas_macros}
\usepackage{graphicx}
\usepackage{epstopdf}
\usepackage{mathrsfs}
\usepackage{hyperref}
\usepackage{amsmath}
\usepackage{xspace}
\usepackage{natbib}
\usepackage{natbib}

\newcommand{\nstars}{83\xspace}
\begin{document}
\title{Discovery of low-metallicity stars in the central parsec of the Milky Way}
\author{Tuan Do\altaffilmark{1,2,7}, Wolfgang Kerzendorf\altaffilmark{3,4,8}, Nathan Winsor\altaffilmark{1,5}, Morten St\o stad\altaffilmark{3}, Mark R. Morris\altaffilmark{2}, Jessica R. Lu\altaffilmark{6}, Andrea M. Ghez\altaffilmark{2}}

\altaffiltext{1}{Dunlap Institute for Astronomy and Astrophysics,
University of Toronto, 50 St. George Street, Toronto M5S 3H4, ON, Canada}
\altaffiltext{2}{UCLA Galactic Center Group, Physics and Astronomy Department,
UCLA, Los Angeles, CA 90095-1547, tdo@astro.ucla.edu}
\altaffiltext{3}{{Department of Astronomy and Astrophysics,
University of Toronto, 50 St. George Street, Toronto M5S 3H4, ON, Canada}}
\altaffiltext{4}{{ESO, Garching, Germany}}
\altaffiltext{5}{Grenfell Campus - Memorial University of Newfoundland, 20 University Drive
Corner Brook, NL, A2H 5G4, Canada}
\altaffiltext{6}{Institute for Astronomy, University of Hawaii}
\altaffiltext{7}{Dunlap Fellow}
\altaffiltext{8}{ESO Fellow}

\begin{abstract}
We present a metallicity analysis of \nstars late-type giants within the central 1 pc of the Milky Way. K-band spectroscopy of these stars were obtained with the medium-spectral resolution integral-field spectrograph NIFS on Gemini North using laser-guide star adaptive optics. Using spectral template fitting with the MARCS synthetic spectral grid, we find that there is large variation in metallicity, with stars ranging from [M/H] $<$ -1.0 to above solar metallicity. About 6\% of the stars have [M/H] $<$ -0.5. This result is in contrast to previous observations, with smaller samples, that show stars at the Galactic center have approximately solar metallicity with only small variations.  Our current measurement uncertainties are dominated by systematics in the model, especially at [M/H] $>$ 0, where there are stellar lines not represented in the model. However, the conclusion that there are low metallicity stars, as well as large variations in metallicity is robust. The metallicity may be an indicator of the origin of these stars. The low-metallicity population is consistent with that of globular clusters in the Milky Way, but their small fraction likely means that globular cluster infall is not the dominant mechanism for forming the Milky Way nuclear star cluster. The majority of stars are at or above solar metallicity, which suggests they were formed closer to the Galactic center or from the disk. In addition, our results indicate that it will be important for star formation history analyses using red giants at the Galactic center to consider the effect of varying metallicity.
\end{abstract}


\section{Introduction}

The metallicity of stars and stellar populations is an important property that allows us to understand their formation and subsequent evolution. Metallicity can also serve as a signature for separating multiple populations of stars formed at different times. When averaged among many stars, metallicity can be used to trace star formation within and between galaxies. 

Chemical abundance measurements of stars in the Milky Way have shown that there is a strong gradient in metallicity  \citep{2013NewAR..57...80F}. The metallicity increases from below solar metallicity in the outskirts of the Milky Way disk to above solar metallicity within the central 5 kpc \citep{2011MNRAS.417..698L}. While the sample of stars in the Milky Way with abundance measurements has increased dramatically with spectroscopic surveys such as APOGEE \citep{2014ApJ...796...38N}, there are abundance measurements of only about a dozen stars in the central 10 pc of the Galaxy \citep{2000ApJ...530..307C,2000AJ....120..833R,2007ApJ...669.1011C,2014arXiv1409.2515R}. These measurements are consistent with the Galactic trend, with a mean [M/H] = $0.14\pm0.06$, and a dispersion of 0.16 dex \citep{2007ApJ...669.1011C}. 

The metallicity measurements of stars within the central 10 pc of the Galaxy is important, because they form the basis of our interpretation of the formation and properties of the Milk Way nuclear star cluster. This cluster is the most massive ($10^7$ $M_\odot$) in the Galaxy and provides us with a template for understanding the nuclei of other galaxies \citep{2009A&A...502...91S,2015MNRAS.447..952C}. The metallicity measurement helps place the cluster in context with the rest of the Galaxy, and serves as a starting assumption when inferring the star formation history and initial mass function (IMF) from the infrared luminosity function \citep{2007ApJ...669.1024M,2011ApJ...741..108P,2013ApJ...764..155L}. It is therefore important to obtain larger spectroscopic samples of stars in this region to obtain robust measurements their physical properties like [M/H]. 

In this study, we combine high angular resolution spectroscopy of red giants with spectral template fitting to directly constrain their effective temperature, gravity, and [M/H]. With adaptive optics (AO) spectroscopy, we are able to increase both the depth and the number of stars with measured [M/H] in this region by a factor of 8. This increase in sample size has revealed a number of stars with significantly lower metallicity than have been previously measured in this region. We also discuss the implications of these measurements. 

\section{Observations}

The spectra in this paper were obtained using the Gemini North Near-Infrared Facility Spectrograph (NIFS) with the natural and laser guide star adaptive optics system ALTAIR. The spectra were obtained using the K broad-band filter (1.99 - 2.40 $\micron$; GN-2012A-Q-41 and GN-2014A-Q-71, PI: Do). The observations span a projected radius of 8 to 22 arcseconds (0.3 to 0.9 pc) from Sgr A*. More details about the observations and data reduction were presented by \citep{2015arXiv150407239S}. We restrict our analysis here to stars with signal-to-noise ratios (SNR) greater than 35 in order to better utilize weak spectral lines, and obtain results that are less sensitive to priors on the parameters. We also consider only late-type stars (F-type or later) with temperatures between 2500-7000 K. In total, we analyze \nstars stars. 

\section{Spectral template fitting}

We fit the observed spectra to a MARCS grid of synthetic models to obtain physical parameters \citep{2008A&A...486..951G}. The MARCS spectral grid spans a range of effective temperature ($T_{eff}$) between 2500 to 7000 K, surface gravity ($\log g$) between -0.5 and 5.0 dex, and scaled solar metallicity (henceforth described by [M/H]) between -4.0 and 1.0 dex. The MARCS grid available online\footnote{\url{http://marcs.astro.uu.se/index.php}} are sampled at intervals of $~100$ K in $T_{eff}$. [M/H] is sampled variable steps from 0.25 dex between $-1 < [M/H] < 1.0$, 0.5 dex between $-3 < [M/H] < -1$. $\log g$ is sampled in steps of 0.5 dex. We consider models of solar composition from \citet{2007SSRv..130..105G} for this analysis. In order to obtain spectra from intermediate grid parameters, we use a linear interpolation between spectra of neighboring grid points. The MARCS grid (R = 20000) is convolved to R = 5400 for comparison with the NIFS spectra. The spectral resolution is determined using the average of the full-width half-maximum of isolated OH sky-lines. The spectral resolution vary by about 10\% between different sky lines though not systematically with wavelength. This amount of variation does not significantly affect the fitted parameters. We ignore stellar rotation in our fit because red giants are observed to have rotational velocities below 10 km s$^{-1}$ \citep{1989ApJ...347.1021G}, which is not resolvable at the spectral resolution of NIFS. We also limit our fitting range to 2.1 to 2.291 $\micron$ to exclude CO lines, which tend to bias our fits (see Section \ref{sec:standards}).

We utilize the Bayesian sampler MultiNest \citep{2009MNRAS.398.1601F,2014A&A...564A.125B} to fit the observed spectra. The fit is done by computing the posterior:
\begin{equation}
P(\theta| D) = \frac{P(D|\theta)P(\theta)}{P(D)}
\end{equation}
where $D$ is the observed spectrum, and the model parameters $ \theta = (T_{eff}, ~\log g, ~\textrm{[M/H]}, v_z)$, where $v_z$ is the radial velocity. The priors on the model parameters are $P(\theta)$ and $P(D)$ is the evidence, which acts as the normalization. The combined likelihood for an observed spectrum is:
\begin{equation}
P(D|\theta) = \prod^{\lambda_n}_{\lambda=\lambda_0}\frac{1}{\epsilon_{\lambda,obs}\sqrt{2\pi}}\exp{(-(F_{\lambda,obs} - F_\lambda(\theta))^2/2\epsilon_{\lambda,obs}^{2})},
\end{equation}
where $F_{\lambda,obs}$ is the observed spectrum, $F_{\lambda}(\theta)$ is the model spectrum evaluated with a given set of model parameters, and $\epsilon_{\lambda,obs}$ is the 1 $\sigma$ uncertainty for each observed flux point. This likelihood assumes that the uncertainty for each flux point is approximately Gaussian. For computational efficiency, we use the log-likelihood in place of the likelihood:
\begin{equation}
\ln P(D|\theta) \propto -\frac{1}{2} \sum^{\lambda_n}_{\lambda=\lambda_0}((F_{\lambda,obs} - F_\lambda(\theta))^2/\epsilon_{\lambda,obs}^{2}).
\end{equation}
We choose to use the MultiNest sampler because the parameter space is often multi-modal, and we find that other techniques like Markov-Chain Monte Carlo are less efficient and often converge at a local maximum instead of the global best-fit solution.

In addition, for each model evaluation, we apply a least-squares fit for a fourth order polynomial that minimizes the difference the observed and model spectrum. In this way, the stellar absorption lines are accounted for in matching the continuum. This procedure however removes our ability to constrain the effective temperature using the shape of the spectrum.  

\begin{figure*}[t]
\center
\includegraphics[width=6.5in]{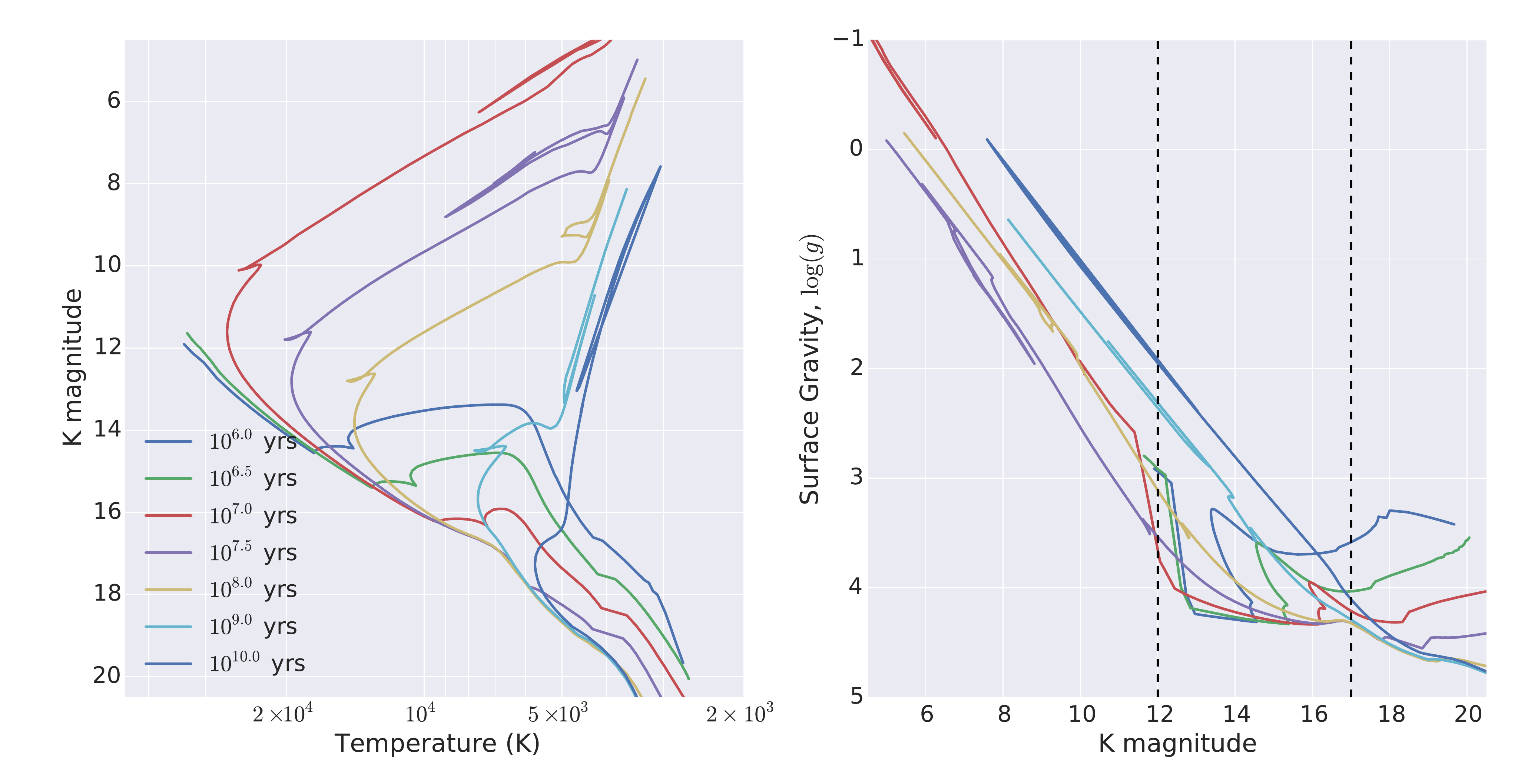}
\caption{\textbf{Left:} PARSEC isochrones \citep{2012MNRAS.427..127B} sampled over the range of observed stellar ages at the Galactic center from the young stars ($< 10$ Myr) to red giants (1-10 Gyr). \textbf{Right:} The surface gravity of stars as a function of K magnitude. We utilize these relationships to limit the range of $\log g$ in the spectral fits. Within the brightness range of most of our measurements (K = 12 - 16 mag), surface gravity varies from $\log g = 4.5$ to $\log g = 2$.}
\label{fig:isochrone}
\end{figure*}

The priors for the model parameters are chosen based on the constraints of the MARCS grid and from stellar evolutionary models for stars that could exist at the Galactic center. The prior on $T_{eff}$ is uniform from 2500 to 7000 K, appropriate for red giants. The prior on [M/H] is uniform from -4.0 to 1.0 dex. To determine the limits on $\log g$, we use the PARSEC isochrones \citep{2012MNRAS.427..127B} for ages between $10^6$ to $10^{10}$ yrs. We then compared the range of $\log g$ with that of the K-band luminosity expected at the Galactic center for these isochrones (Figure \ref{fig:isochrone}). Based on these models, we set the priors on $\log g$: -0.5 $<\log g<$ 4.0 for $K < 12$ mag and 2.0 $<\log g<$ 4.5 for $K \geq 12$ mag. 

\section{Characterizing uncertainties}
\label{sec:uncertainties}
\subsection{Uncertainty from Interpolation}

While interpolation allows us to produce spectra with arbitrary model parameters, it also increases the uncertainty in the model spectrum. In order to characterize this uncertainty, we remove a grid point, compute an interpolated spectrum at that same point, and then fit the stellar parameters of the interpolated spectrum using the original grid. This represents the maximum deviation due to interpolation, because the interpolated spectra used for our analyses will be, at most, about half the distance from a reference grid point. Repeating this process for the entire grid, we find that  $\sigma_{T_{eff}} = 50$ K, $\sigma_{[M/H]} = 0.1$, and $\sigma_{\log g} = 0.1$, with no systematic offsets in the fitted parameters. We therefore include the interpolation uncertainty by adding these values in quadrature with the statistical uncertainty for the stellar parameters. These uncertainties are larger than the statistical uncertainties, but are small compared to model uncertainties.

\subsection{Fit comparison with standard spectra}
\label{sec:standards}
In order to assess possible model uncertainties in the fits, we derive the physical parameters of stars from the SPEX stellar spectral library for comparison to previous measurements \citep{
  2009ApJS..185..289R}. While this library has lower spectral resolution (R = 2000) than NIFS, it is the most complete publicly available spectral library in the K-band spanning the range of parameter space of our sample. Importantly, many of the SPEX spectra have been observed previously and have had their stellar parameters measured \citep[tabulated by][]{2013A&A...549A.129C}. By comparing to these previous values, we can estimate the range of model uncertainties. 
  
We include constraints on $\log g$ with knowledge of the luminosity of the stars, similar to our analysis of the Galactic center stars. Stars of luminosity class III were limited to $2 < \log g < 4$. Stars with luminosity class I (supergiants) and II were limited to $-0.5 < \log g < 2$. Luminosity class V (main sequence) were limited to $3 < \log g < 5.5$. We also limit the wavelength range of the fit from 2.1 to 2.291 $\micron$, as we find there are significant biases in $\log g$ and [M/H] when including CO lines. When compared to the values from \citet{2013A&A...549A.129C}, the mean and standard deviation of the fit residuals are: $\Delta_{[M/H]} = -0.2$, $\sigma_{[M/H]} = 0.3$, $\Delta_{\log g} = 1.0$, $\sigma_{\log g} = 0.9$, $\Delta_{T_{eff}} = 50$ K, and $\sigma_{T_{eff}} = 400$. Figure \ref{fig:spex} shows the correlation between our fits to that of the literature, and some examples of SPEX spectra compared to our sample with similar [M/H]. The offsets from the reference measurements are likely due to systematics in the MARCS model spectra, and the fact that the previous measurements were made by different authors, which can differ often by 0.2 dex and, in some cases, up to 1 dex in [M/H] \citep{2013A&A...549A.129C}. Observation of these spectral standards with NIFS will be useful to further quantify the comparison between different abundance measurement techniques. We include these rms values in the uncertainty for our measured parameters by adding them in quadrature with the other uncertainties. These systematic uncertainties dominate over all other sources. The comparison with previous measurements also shows that the $\log g$ values are prone to biases, but the values for [M/H] and $T_{eff}$ are consistent to the level of 0.3 dex and 400 K, respectively, which are accurate enough to obtain significant constraints on [M/H] and $T_{eff}$ of stars at the Galactic center. 

\begin{figure}[]
\center
\includegraphics[width=\linewidth]{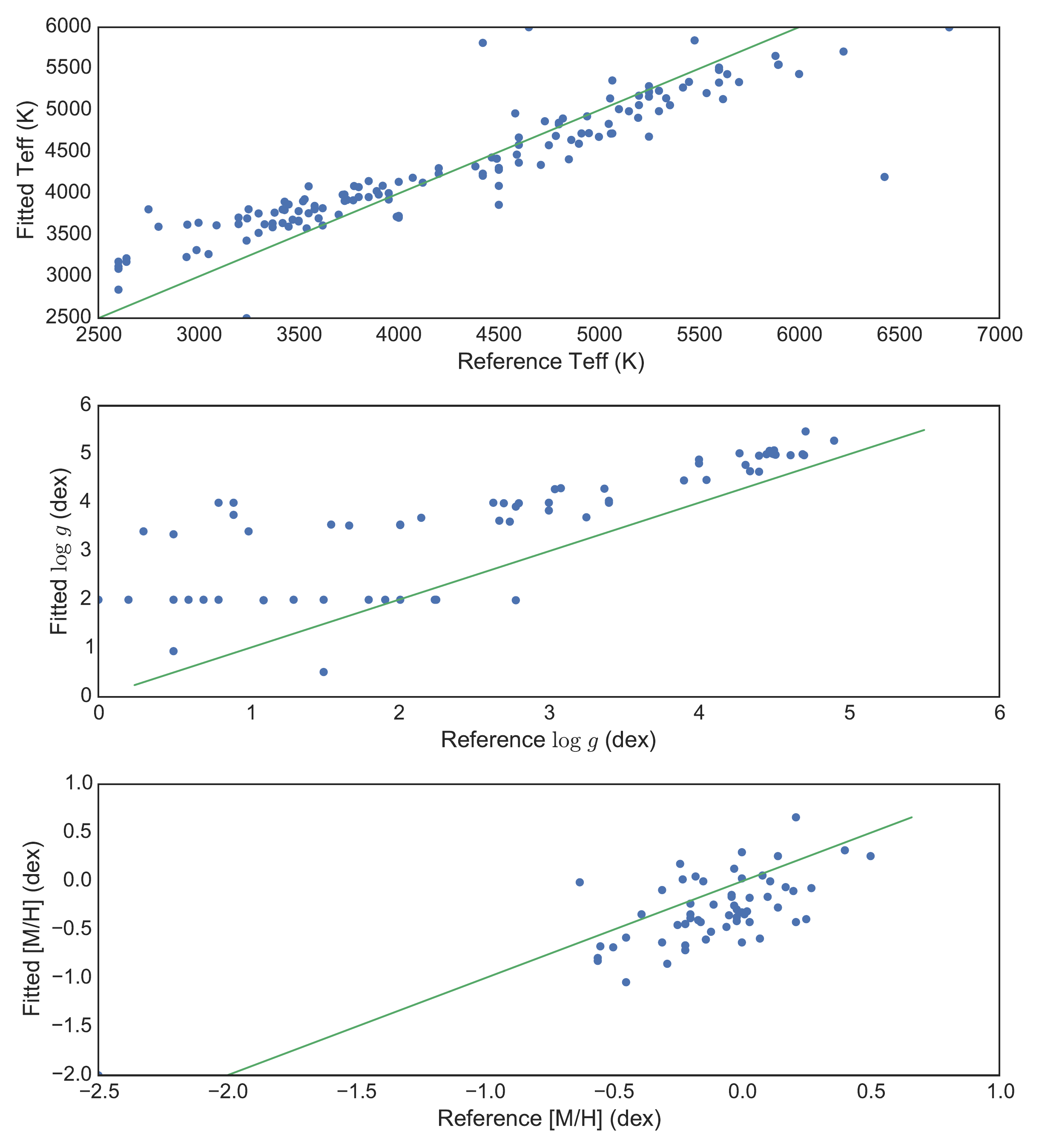}
\caption{Comparison of the best-fit values for the SPEX spectral library to literature values tabulated in \citet{2013A&A...549A.129C} for: (top) $T_{eff}$, (middle) $\log g$, (bottom) [M/H]. More stars from SPEX have $T_{eff}$ measurements than $\log g$ or [M/H]. }
\label{fig:spex}
\end{figure}

\begin{figure*}[]
\center
\includegraphics[width=7in]{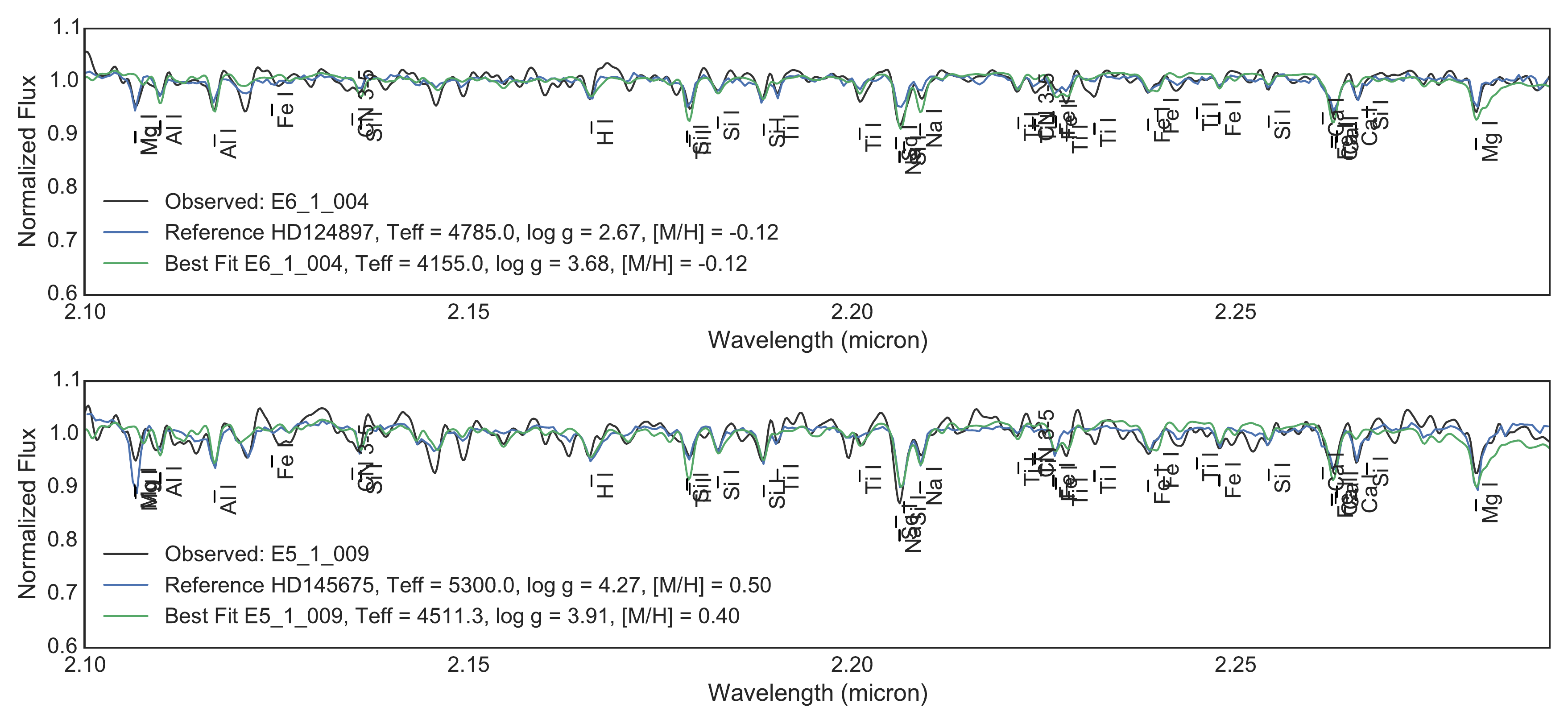}
\caption{Examples of stars in our sample convolved to the spectral resolution of R = 2000 (black) to compare with a spectrum from the SPEX library with a similar [M/H] (blue). The best fit MARCS model (blue) is able to fit most spectral features. Some features, such as Mg I near 2.10 $\micron$ and Ca I near 2.27 $\micron$ are not well reproduced by the model. In general, the fits using the MARCS grid are consistent with previous measurements of [M/H] for the SPEX library with a dispersion of about 0.3 dex (see Section \ref{sec:standards})}
\label{fig:spex}
\end{figure*}

\subsection{Comparison to the solar spectrum}
\label{sec:solar}
\begin{figure*}[]
\center
\includegraphics[width=\linewidth]{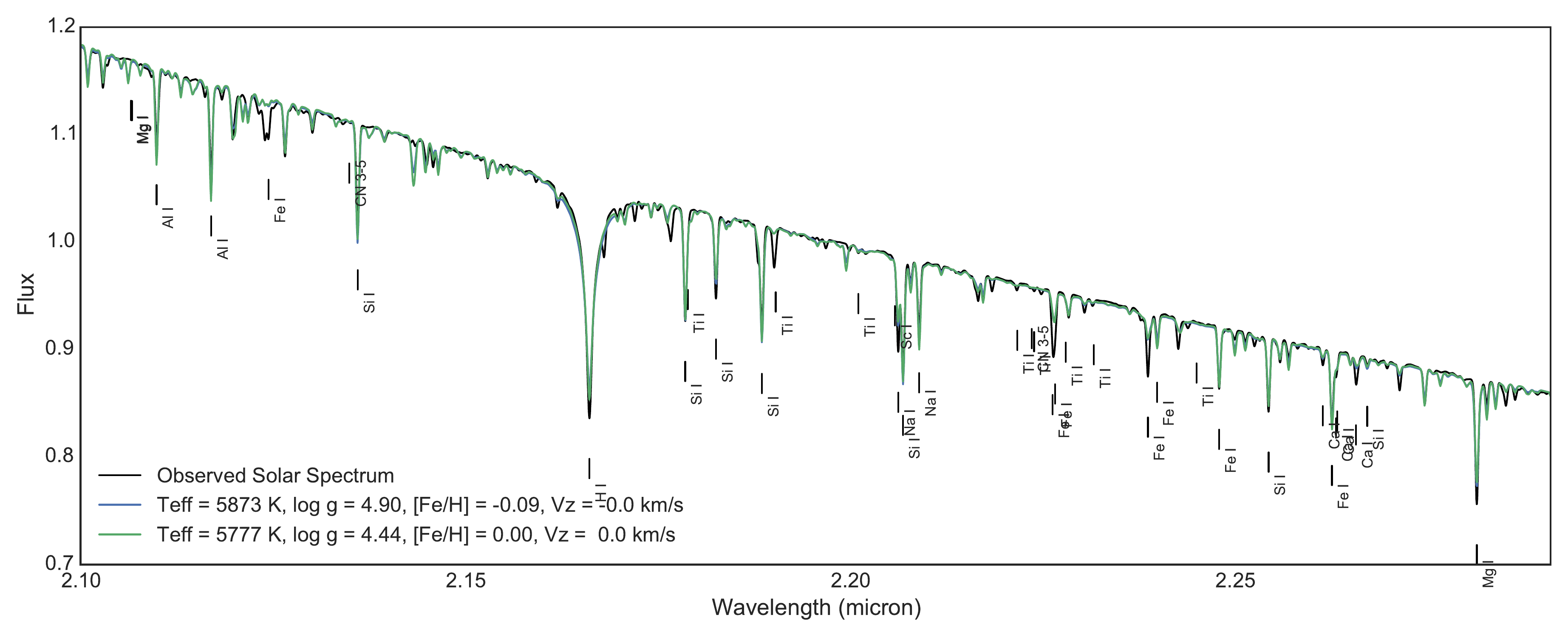}
\caption{The best fit MARCS model (blue) compared to the observed solar spectrum in the near-infrared  (black; NSO/NOAO). The fiducial solar model is also shown (green). In general most of the lines are well fit, with the best fit within the range of uncertainties we find from comparison to the SPEX spectral templates. A few lines have are not well matched by the model, which likely is a source of systematic uncertainty in our fits.}
\label{fig:solar}
\end{figure*}

We also test our model and fitting method with the spectrum of the sun and find a consistent fit (Figure \ref{fig:solar}). The solar spectrum\footnote{Obtained from \url{http://www.eso.org/sci/facilities/paranal/decommissioned/isaac/tools/spectroscopic_standards.html} using NSO/Kitt Peak FTS data produced by NSF/NOAO.} was convolved to a spectral resolution of R = 5400 and fit with the same wavelength range as used to fit the NIFS and SPEX spectra (2.1 to 2.291 $\micron$), with uniform priors, $T_{eff} = 2500$ to 7000 K, $\log g = -0.5$ to 5.0, [M/H] = -2.0 to 1.0, $v_z = -600$ to 600 km s$^{-1}$. The best fit values are: $T_{eff} = 5872$ K, $\log g = 4.9$, [M/H] = -0.089, $v_z = 0.032$ km s$^{-1}$. For comparison, the fiducial values for the sun used by the MARCS grid \citep{2008A&A...486..951G} are: $T_{eff} = 5777$ K, $\log g = 4.44$, [M/H] = 0.00, $v_z = 0$ km s$^{-1}$. The these values are consistent with the fit given the uncertainty from interpolation and estimates of the systematic uncertainty derived in Section \ref{sec:standards}. While most of absorption lines are well matched, there are a few that are not well matched that may be the source of the small systematic offsets from the fiducial values. These lines include the Fe I at 2.22632 $\micron$, 2.22662 $\micron$, and 2.23990 $\micron$, and Ca I at 2.26574 $\micron$. However, excluding spectral features with the large deviations have negligible impact on the fitted parameters, as there are many spectral features that are incorporated in the global fit (see Appendix for more details). 

\section{Results}

\begin{figure*}[bt]
\center
\includegraphics[width=5in]{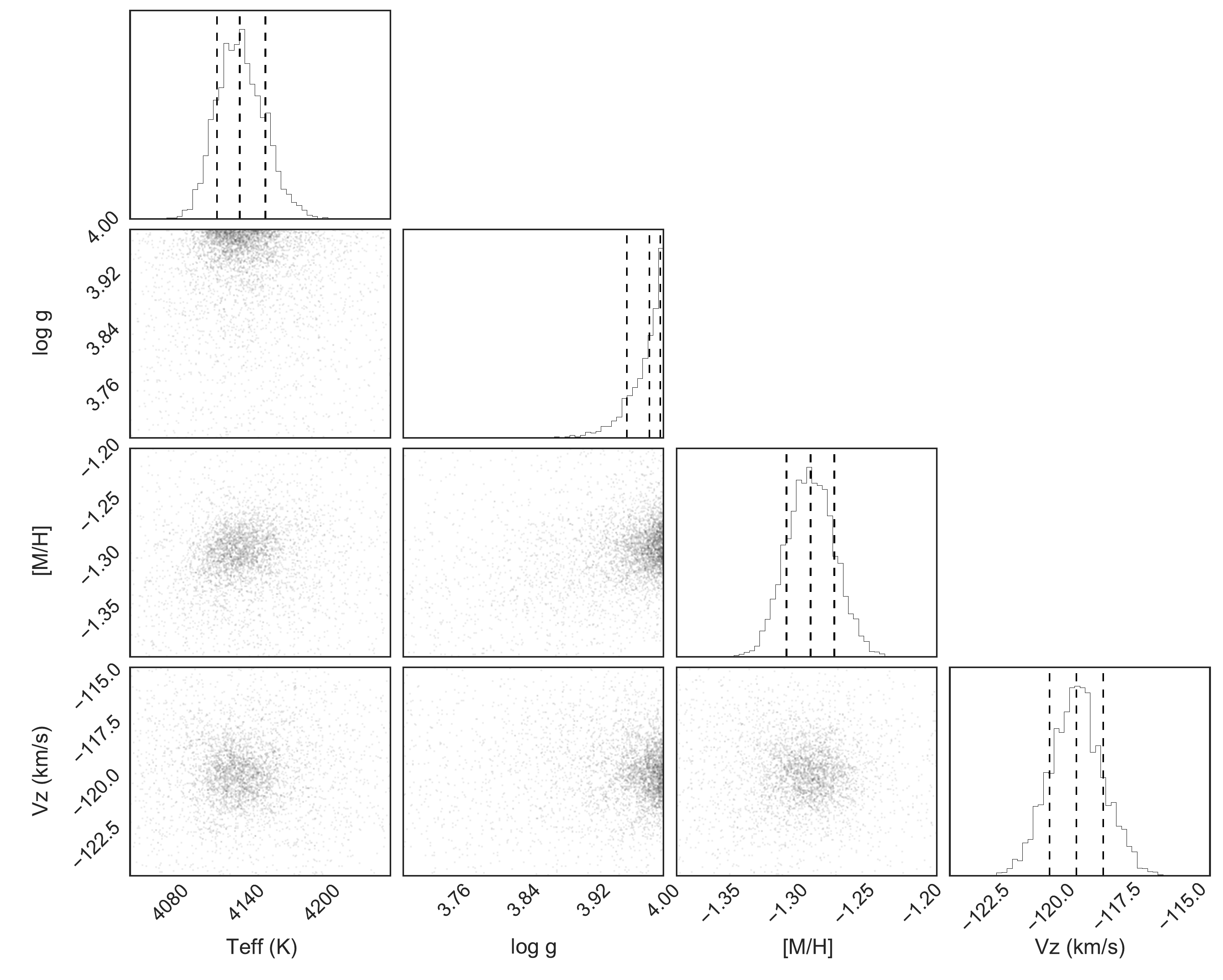}
\caption{The joint probability distributions for T$_{eff}$, $\log g$, [M/H], and $v_z$, as well as their 1-dimensional probability distributions for the stellar parameters for NE1-1 003.  }
\label{fig:pdfs}
\end{figure*}

For each of the \nstars spectra, we fit for the 4 physical parameters: $T_{eff}$, $\log g$, [M/H], and $v_z$. We report the central value of the probability distribution for each of these parameters marginalized over all other model parameters in Table 1. For most stars in the sample, there are correlations between $T_{eff}$, $\log g$, and [M/H], depending on the star, which emphasizes the necessity to fit them simultaneously. As expected, there is no correlation between $v_z$ and the other 3 parameters.  Figure \ref{fig:pdfs} shows an example of the joint probability distribution functions (PDFs) for the 4 stellar parameters in the fit for the star NE1-1 003. These PDFs show that the statistical uncertainties are generally very small compared to the systematic uncertainties estimated in Section \ref{sec:uncertainties}. 

\begin{figure*}[bt]
\center
\includegraphics[width=6.5in]{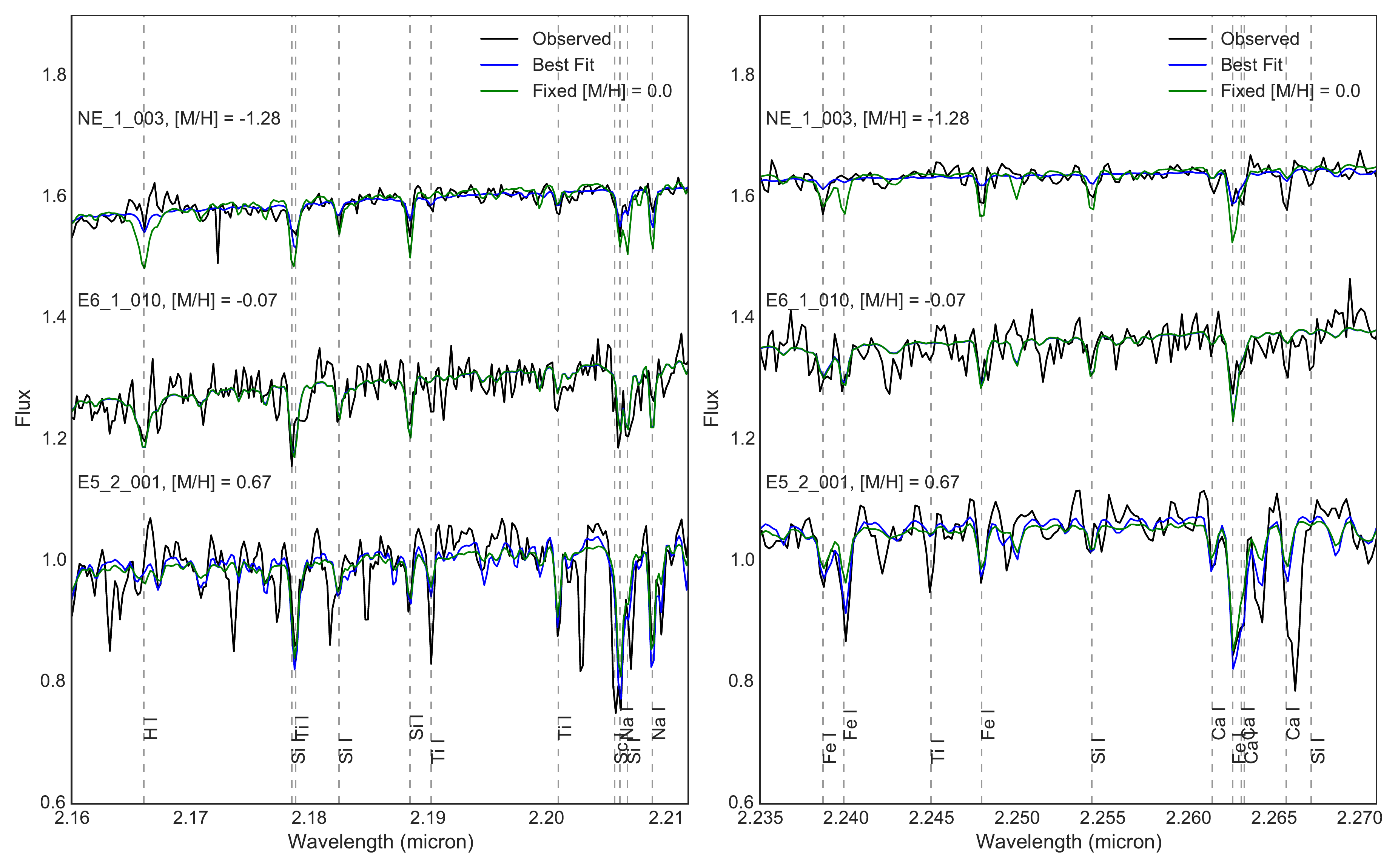}
\caption{Examples of spectra with a range of [M/H] along with their best fits (blue) compared to a fit with [M/H] fixed to 0.0. Labeled are the best fit [M/H] values. The three stars show examples of metal-poor, solar metallicity, and super-solar metallicity stars. These two wavelength regions of K-band were chosen to illustrate the combination of temperature-sensitive lines such as H I compared to metallicity sensitive lines such as Fe I and Si I. The MARCS spectra appear to a good match at solar or below solar metallicities. At high metallicities, the model is unable to reproduce some of the observed lines, likely leading to larger systematic uncertainties in the value of [M/H] for these stars.}
\label{fig:fit_multi}
\end{figure*}

\begin{figure*}[bt]
\center
\includegraphics[width=6in]{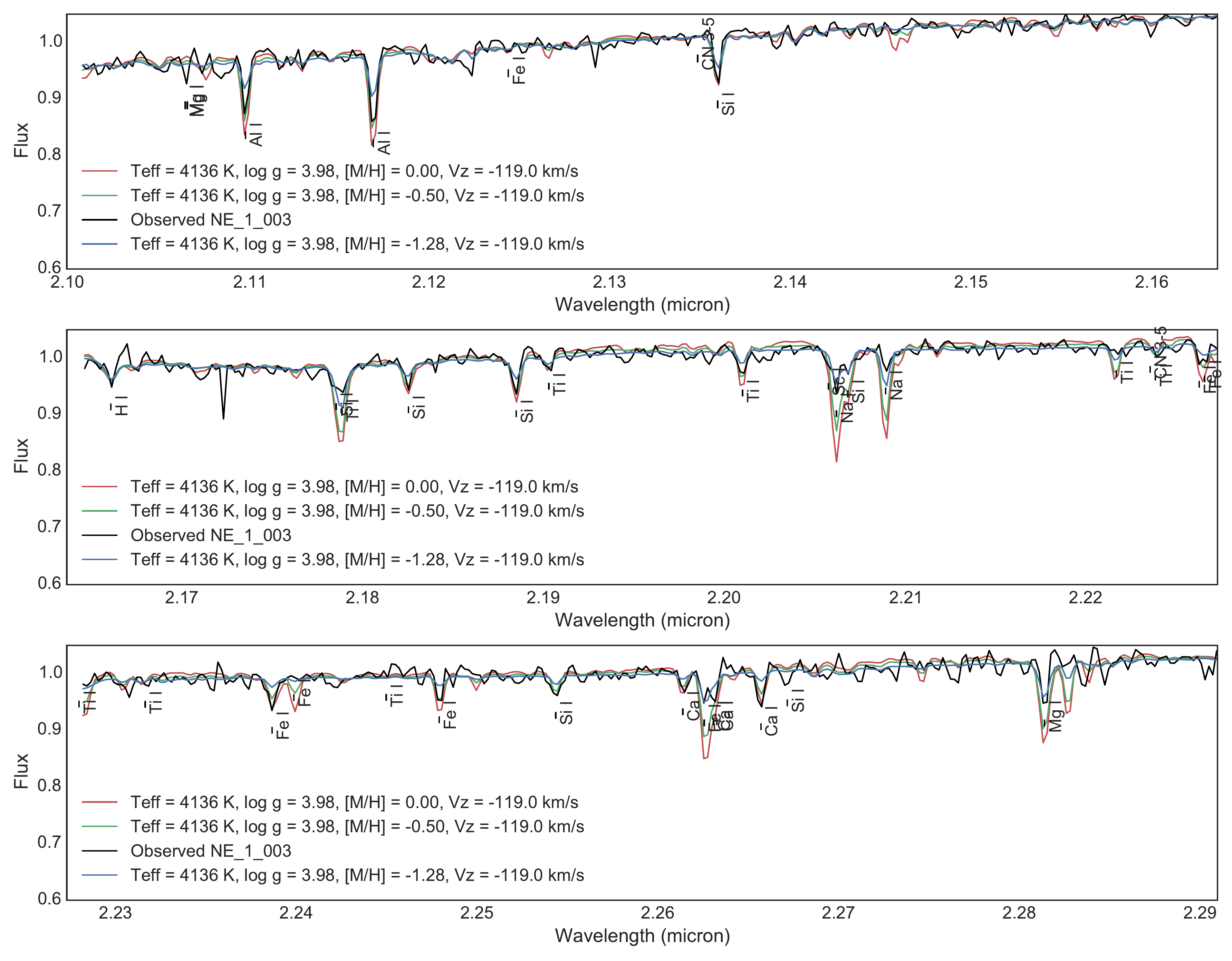}
\caption{Comparison of the best fit model for NE1-1 003 ([M/H] = -1.28) to [M/H] = -0.5 and 0.0 to show the effect on the spectrum with increasing metallicity.}
\label{fig:example_compare}
\end{figure*}

\begin{figure}[]
\center
\includegraphics[width=\linewidth]{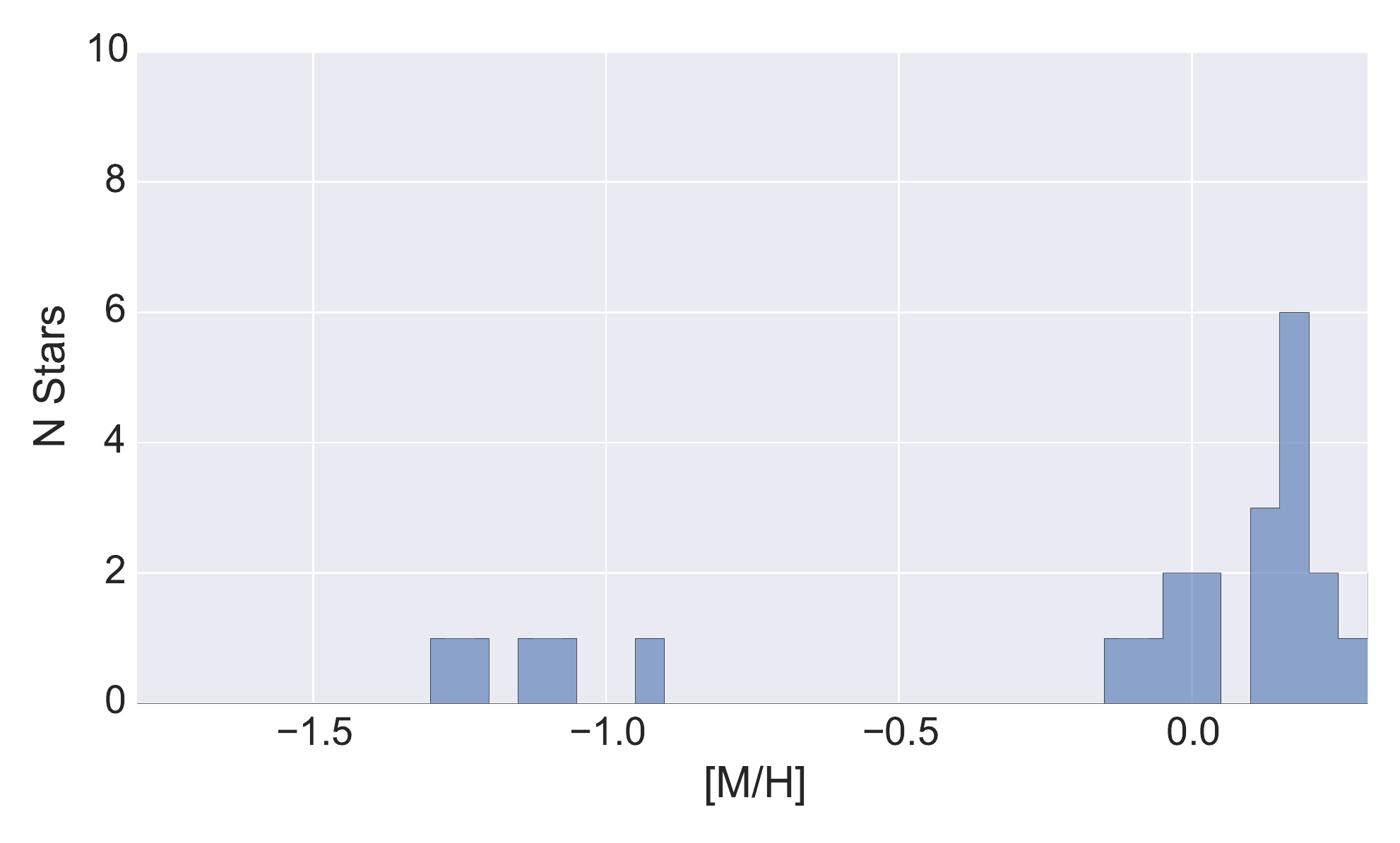}
\caption{Distribution of [M/H] measurements for stars with [M/H] $<$ 0.3 (remaining stars have higher measured [M/H], but are likely subject to larger systematic uncertainties).}
\label{fig:hist}
\end{figure}

The fits to the physical parameters reveal the existence of stars with low metallicity at the Galactic center. These stars have unusual spectra compared to the rest of the sample, showing low CO, Na, and Ca equivalent widths \citep[e.g.][]{2011ApJ...741..108P,2013ApJ...764..154D}. In Figure \ref{fig:fit_multi}, we show selected stars with a range of metallicity, from low to super-solar. For comparison, we also show the best-fit spectra when the metallicity is fixed to [M/H] = 0.0. Certain lines such as Si I and Fe I are sensitive to [M/H], while others, especially H I at 2.1661 $\mu m$, are more sensitive to temperature. For example, the star NE1-1 003 is better fit with low metallicity of [M/H] = -1.28 and $T_{eff}$ = 4136 K, compared to one with fixed [M/H] = 0.0, and best fit $T_{eff} = 5157$ K (Figure \ref{fig:fit_multi}). NE1-1 003 lacks the stronger Br $\gamma$ line at 2.1661 $\mu$m that would be required for the $T_{eff} = 5157$ K fit, and contains weak Fe I and Si I lines that are more consistent with low [M/H]. Some of the lines that show mismatches between the model and observations, such as Fe I at 2.23990 $\micron$, and the Ca I line at 2.26574 $\micron$ (Figure \ref{fig:fit_multi} and \ref{fig:example_compare}). These lines are also mismatched in our solar spectrum comparison (Section \ref{sec:solar}). These mismatches may be a concern if deriving individual elemental abundances, but as our fit aims to determine relative scaled solar abundances, the impact of these mismatch is lessen, as verified by our comparisons in Section \ref{sec:standards} and \ref{sec:solar}. While there may be systematic uncertainties in the absolute measurement of [M/H], the conclusion that these stars must be low metallicity compared to most of the sample is robust. In total, 5 of \nstars (6\%) of the stars have [M/H] $< -0.5$ (Figure \ref{fig:hist}).

Most of the stars have higher than solar metallicity, with a sample mean of [M/H] = 0.4 and a standard deviation of 0.4 dex. This suggests that there are many stars with super-solar metallicity at the Galactic center, though they are likely subject to greater systematic uncertainties. The MARCS models are unable to reproduce many features in the metal-rich sample. For example, in Figure \ref{fig:fit_multi}, the star E5-2 001 shows strong features at 2.1898 and 2.2653 $\micron$, which are not present in the MARCS models. This may be due to differences from solar composition or missing atomic and molecular lines in generating the spectra. In addition, at medium spectral-resolution, the difference between solar-metallicity and high-metallicity spectra are much smaller than that between solar-metallicity and low-metallicity sources (Figure \ref{fig:fit_multi}).  In order to resolve these issues, it will be necessary to expand the comparison to non-solar composition models, and to obtain a larger sample of high spectral resolution observations to calibrate these [M/H] measurements.  While the large spread in [M/H] is likely robust as they are relative measurements, conclusions relying on the location of the peak or shape of the metallicity distribution are probably unreliable at this time. 

\section{Discussion}

\subsection{Foreground/background sources and spectroscopic contamination}

Interpretation of the results of this study depends in part on whether the sources belong to the nuclear star cluster, or are foreground or background sources. Some obvious foreground sources can be excluded based on their blue colors,  and are not in the present sample \citep{2015arXiv150407239S}. The remaining foreground/background contaminants are likely stars from the inner bulge of the Milky Way. The most recent estimate of the number density of stars in the inner bulge in the near-infrared was made by \citet{2012ApJ...751..132C} with their measurement of the proper motion of the Arches star cluster, about 26 pc in projection from the Galactic center. Using their proper motion identification of field stars, we find that the likely number of foreground/background sources is 0.08 stars/arcsec$^{-2}$ with $K < 15.5$. With a coverage of about 99 arcsec$^{2}$ in the current survey, we expect about 8 stars to be from the inner bulge. It is unlikely that the low-metallicity stars found in this study belong to the inner bulge, as the inner bulge was measured to have [M/H] = $-0.16\pm0.12$ \citep{2012ApJ...746...59R,2014arXiv1409.2515R}. 

A second source contamination, especially at low-metallicity, may be Milky Way halo interlopers. Using the Besan\c{c}on Model \citep{2003A&A...409..523R}, we estimated the expected number of halo stars expected between 10 -- 18\,mag in K (including extinction) using an extinction law of $A_V=3.5$mag/kpc. Within our field of view, the model predicts 0.0004 halo stars, which is negligible. 

The presence of a nearby companion would affect the spectra of the stars, biasing the measured stellar parameters. Early-type stellar (B-type or earlier) companions will introduce the strongest bias. These companions may either be physically close binaries or projected pairs. However, this is unlikely for two reasons: (1) there are very few ($\sim2$) early-type stars in the region of this survey \citep{2015arXiv150407239S}, so chance superposition is negligible, (2) physical binaries of a late-type giant and an early-type star are unlikely based on stellar evolution, as the early-type star will have a lifetime of $< 100$ Myr, while the late-type giant is closer to 1 Gyr in age.

\subsection{Comparison with previous work}

There have been few measurements of individual [M/H] values for stars in the central parsec. About a dozen bright late-type stars, most of which are red supergiants or AGB stars have been measured at high spectral resolution. \citet{2000ApJ...530..307C} was the first to measure the metallicity of the red supergiant, IRS 7 ([M/H] = -0.02 $\pm 0.13$), located about 5.5$\arcsec$ (0.22 pc) from the Galactic center. Subsequently, \citet{2000AJ....120..833R} measured 9 red supergiants and AGB stars (K $< 8$ mag) within 2.5 pc of the Galactic center and found a mean [M/H] = 0.12$\pm0.22$. The [M/H] of their sample range from -0.29 to 0.49, with typical uncertainties of 0.3 dex. \citet{2007ApJ...669.1011C} re-observed 5 of the same sources and found a mean metallicity of [M/H] = 0.14, and a dispersion of 0.16 dex. More recently, \citet{2014arXiv1409.2515R} observed 9 fainter giants with K $< 12$ mag within 2.5$\arcmin$ and 3.5$\arcmin$ (6 to 8.4 pc) from the Galactic center. These stars have [M/H] = 0.11$\pm0.15$, similar to the observations in the central parsec.

The measured range of [M/H] in our current sample is large compared to previous measurements. The value of [M/H] for our sample of \nstars stars range from $< -1.0$ to $> 0.8$ dex. This wide range may be the result of the larger sample of stars, which allows us to detect the rarer low-metallicity stars. Our sample does not overlap the previous ones, and the stars are also significantly fainter than the previous measurements with the bulk of the sample between $12 < K < 15.5$ mag. These stars are thus closer to the red clump than the more evolved red-supergiants and AGB stars previously observed in this region. In addition, there may be systematic effects introduced by our medium-resolution spectra compared to the previous high-spectral resolution observations. The method of deriving [M/H] also vary between the different studies. For example,\citep{2000AJ....120..833R} used the MOOG spectral synthesis code \citep{1973PhDT.......180S} while\citet{2014arXiv1409.2515R} used \textit{Spectroscopy Made Easy} \citep{1996A&AS..118..595V}. In order to characterize these uncertainties, it will be useful to obtain compare these methods for stars at the Galactic center. If confirmed, some of these stars may be the most metal-rich ever found in the Galaxy. 

\subsection{Implication for the formation of the nuclear star cluster}

The metallicity of the nuclear star cluster encodes information about its history and initial chemical composition. Until this study, no low-metallicity stars had been found there. They may represent stars that were in this region early in the history of the Milky Way, or may have migrated over time to the Galactic center. Another possibility is that these stars arrived through the infall of globular clusters, which we see today with average [M/H] similar to the low [M/H] stars detected in this study. In fact, one of the prevailing theories for nuclear star cluster formation is that they represent the build up of globular clusters over time \citep[e.g.][]{1975ApJ...196..407T,2008ApJ...681.1136C,2012ApJ...750..111A}. Measurement of the metallicity of these stars offers a new opportunity to test this theory. Globular clusters in the Milky Way generally have [M/H] of about -1.0 to -2.0 \citep{1996AJ....112.1487H}. The highest metallicity clusters are slightly below solar metallicity, such as Terzan 5 at [M/H] = -0.23. If the globular clusters we see today are representative of the past population, the small fraction of low metallicity stars in our sample suggests that globular clusters are only a small contributor to the origin of the Milky Way nuclear star cluster. 

Accounting for the existence of super-solar metallicity stars will be important to determine the origin of the nuclear star cluster, as these represent the majority of stars. The source of these stars may be from the Galactic disk, where the metallicity of stars are generally higher. Super-metal rich stars in the solar neighborhood, with [M/H] up to 0.6, may be from the inner disk or the Galactic bulge \citep{2011A&A...535A..42T,2013A&A...549A.147B,2013NewAR..57...80F}. In order to determine the origin of these stars in the nuclear star cluster, we will need to identify the systematic uncertainties in the model (Figure \ref{fig:fit_multi}), and obtain high resolution spectroscopy to measure stellar abundance ratios. Abundance ratios such as [$\alpha$/Fe], [Ca/Fe], [Si/Fe], etc., can be matched to signatures from other parts of the Galaxy such as the disk, bulge, globular cluster, or local dwarf galaxies. 

\subsection{Implication for stellar population analyses}

\begin{figure}[bt]
\center
\includegraphics[width=\linewidth]{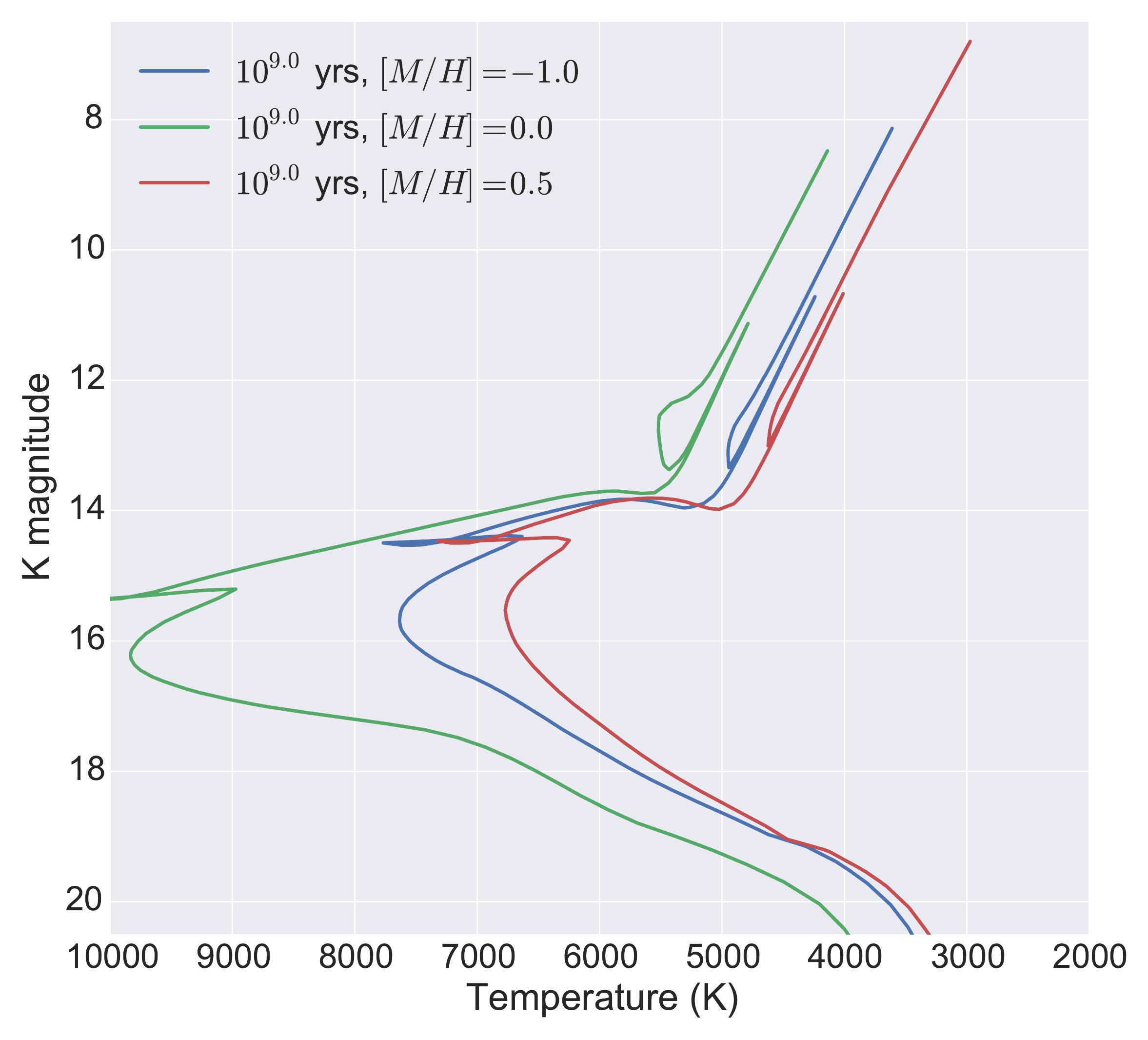}
\caption{Evolutionary tracks from the PARSEC model \citep{2012MNRAS.427..127B} at [M/H] = -1.0, 0, +0.5 at 10$^9$ yrs, showing the difference in temperature and $K$-band magnitude. The $K$-band magnitudes are adjusted for the distance to the Galactic center (8 kpc) and the $K$-band extinction ($A_K$ = 3.0). This shows that interpretation of the stellar population of the Galactic center can change significantly based on its metallicity.}
\label{fig:lum}
\end{figure}

The detection of the large spread of metallicity of stars at the Galactic center has the potential to change the way we study this region. It may be necessary to revisit the measurements of the star formation history and IMF of the nuclear star cluster \citep{2007ApJ...669.1024M,2011ApJ...741..108P,2013ApJ...764..155L}. These studies have so far assumed solar metallicity in their models. These analyses require a translation from the luminosity function into a mass function through the use of evolutionary and atmospheric models, which can strongly depend on the metallicity of the stars. In Figure \ref{fig:lum}, we show a comparison between the $K$-band luminosity function resulting from 0.5 to 2 M$_\odot$ stars of a cluster with an age of 10$^9$ yrs with [M/H] = -1.0, 0.0, and +0.5 using the PARSEC evolutionary tracks \citep{2012MNRAS.427..127B}. These isochrones show that there are significant differences in the luminosity   and temperature of stars at different metallicities, which will need to be accounted for when deriving a star formation history of this region.

\section{Conclusion}

We presented stellar parameter estimates using K-band spectra of \nstars stars within 1 pc of the center of the Galaxy, and we find a significant spread in metallicity of stars, ranging from 10 times below solar to super-solar metallicities (about 6\% of the sample have [M/H] below -0.5). This variation in metallicity shows that the Milky Way nuclear star cluster is not composed of a simple stellar population, which previous work has also shown with spectroscopy and luminosity functions \citep[e.g.][]{2007ApJ...669.1024M,2015MNRAS.447..952C}. Future measurements of the star formation history of this region will need to take the variation in metallicity into consideration. The low-metallicity stars found in this study are consistent with the range of [M/H] observed for globular clusters, which is predicted by the theory that infalling globular clusters contribute to the build up of nuclear star clusters \citep{2008ApJ...681.1136C,2012ApJ...750..111A}. This fraction however is small, and thus this mechanism is not likely to be the source of the Milky Way nuclear star cluster. The evidence for super-solar metallicity, on the other hand, points towards a significant contribution from stars with origins near the center of the Galaxy \citep{2013NewAR..57...80F}.

This paper uses data from Gemini observations GN-2012A-Q-41 and GN-2014A-Q-71. We wish to thank the anonymous referee and Norm Murray for useful comments.

\appendix

\section{Testing the effect of mismatched spectral features}
\label{sec:resdiuals}
\begin{figure*}[bt]
\center
\includegraphics[width=\linewidth]{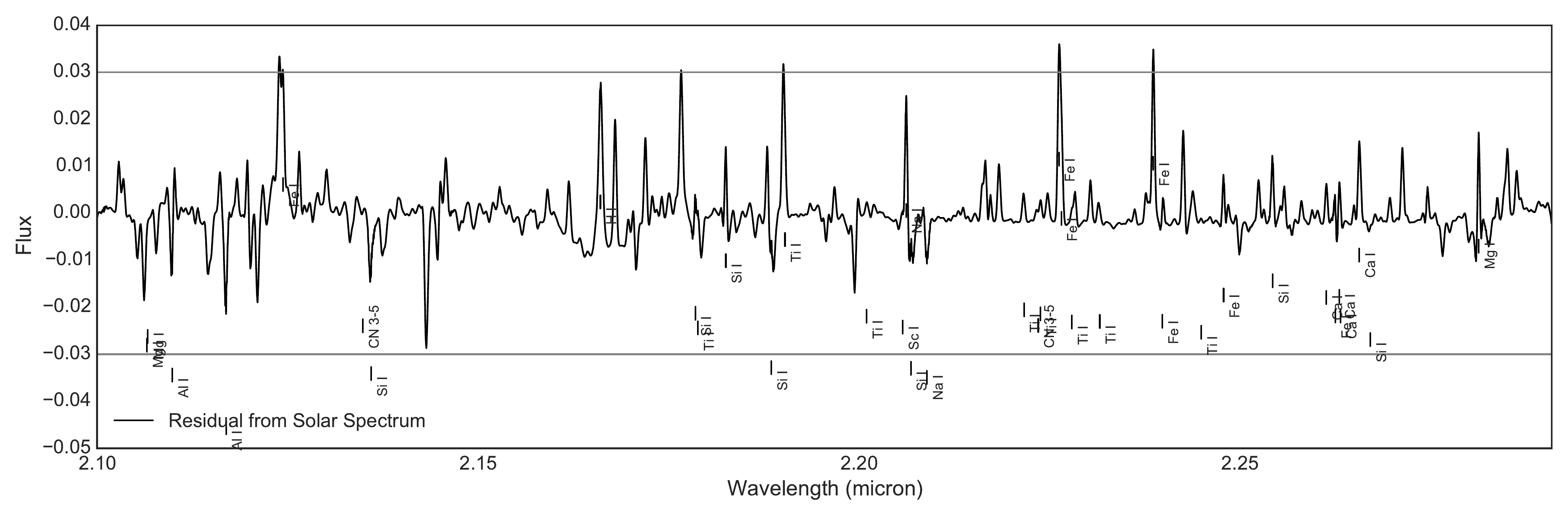}
\caption{Residual of the best fit MARCS model to the Solar spectrum. In order to test the robustness of our fit against large deviations in the model, we exclude regions where the residuals are greater than 3\%. Excluding these regions do not significantly impact the fits for the solar parameters ([M/H] changed by less than 0.02 dex, an order of magnitude smaller than the uncertainties for the stars in our sample).}
\label{fig:solar_residual}
\end{figure*}

Here we analyze possible effects on the stellar parameter fits from mismatches of some spectral features from the model spectra. For example, in the comparison with the Solar spectrum (Figure \ref{fig:solar}), there are a few lines, such as the Fe I at 2.22632 $\micron$, 2.22662 $\micron$ that are not well represented in the model. To determine the effect of this mismatch on the fitted parameters, we compute the residuals between the best fit model and the Solar spectrum to determine regions that deviate by more than 3\%. 5 features exceed this level of deviation (Figure \ref{fig:solar_residual}). We exclude the 4 spectral channels around wavelength regions at  2.17688, 2.19012, 2.2386.9, 2.2263.4 $\micron$, and 16 spectral channels around 2.12418 $\micron$, and refit the spectrum. The best fit parameters are $T_{eff} = 5902$ K, $\log g = 4.9$, [M/H] = -0.07, $v_z = 0.45$ km s$^{-1}$. These values are consistent with the fit to the full spectrum ($T_{eff} = 5872$ K, $\log g = 4.9$, [M/H] = -0.089, $v_z = 0.032$ km s$^{-1}$). We also repeat the analysis of the SPEX spectral library described in Section \ref{sec:standards} by masking the same features. We find that the mean difference and standard deviation between the sample with clipping and the one without to be: $T_{eff} = -2.6\pm26$ K, $\log g = -0.010 \pm0.14$ dex, [M/H] $= 0.002\pm0.04$ dex. These differences are about an order of magnitude smaller than the statistical uncertainties, so we conclude that these spectral features do not affect the fitted stellar parameters.

\LongTables
\begin{deluxetable*}{cccccccccccr}
   \tablecolumns{12}
   \tablewidth{0pc}
   \centering
   \tabletypesize{\scriptsize}  % make the font smaller so that it will fit
   \tablecaption{Measured physical parameters of stars at the Galactic center}   \tablehead{\colhead{Name\tablenotemark{a}} & \colhead{$K$} & \colhead{SNR} & \colhead{RA Offset ($\arcsec$)} & \colhead{DEC Offset ($\arcsec$)} & \colhead{$T_{eff}$ (K)} & \colhead{$\sigma_{T_{eff}}$\tablenotemark{b} (K)} & \colhead{$\log g$} & \colhead{$\sigma_{\log g}$\tablenotemark{b}} &
\colhead{[M/H]} & \colhead{$\sigma_{[M/H]}$\tablenotemark{b}} & $v_z$ (km s$^{-1}$)}
\startdata
E5-1 001 & 12.0 &  42 & 15.17 & -0.49 & 3497 &  413 & 3.04 & 0.91 & 0.96 & 0.32 &  -54 \\ 
E5-1 002 & 12.6 &  54 & 14.52 &  1.47 & 3671 &  414 & 2.84 & 0.91 & 0.55 & 0.32 &  -64 \\ 
E5-1 003 & 13.2 &  44 & 15.00 &  1.06 & 3597 &  414 & 3.09 & 0.91 & 0.85 & 0.32 &  -58 \\ 
E5-1 006 & 14.7 &  40 & 16.80 &  0.36 & 4076 &  417 & 3.86 & 0.91 & 0.44 & 0.32 &  -29 \\ 
E5-1 007 & 15.4 &  41 & 15.25 & -0.23 & 3754 &  414 & 3.18 & 0.91 & 0.40 & 0.32 &  -57 \\ 
E5-1 008 & 15.1 &  36 & 15.02 &  0.68 & 3915 &  418 & 3.56 & 0.91 & 0.43 & 0.32 & -112 \\ 
E5-1 009 & 14.8 &  39 & 13.83 & -0.29 & 4511 &  419 & 3.91 & 0.91 & 0.40 & 0.32 &  -61 \\ 
E5-1 012 & 15.2 &  36 & 14.12 & -0.04 & 4072 &  420 & 3.64 & 0.91 & 0.63 & 0.32 &  112 \\ 
E5-1 014 & 15.4 &  38 & 16.53 &  1.32 & 4197 &  421 & 3.82 & 0.91 & 0.27 & 0.33 &  -52 \\ 
E5-1 015 & 15.7 &  39 & 14.78 & -0.36 & 4180 &  415 & 3.91 & 0.91 & 0.25 & 0.32 &  -75 \\ 
E5-1 016 & 15.5 &  42 & 15.75 & -0.69 & 4145 &  414 & 3.89 & 0.91 & 0.13 & 0.32 &   58 \\ 
E5-1 026 & 16.2 &  36 & 15.61 &  0.30 & 4310 &  418 & 3.93 & 0.91 & -0.05 & 0.33 & -134 \\ 
E5-1 042 & 14.9 &  35 & 15.28 &  1.05 & 4161 &  417 & 3.82 & 0.91 & 0.56 & 0.32 &  228 \\ 
E5-2 001 & 11.3 &  60 & 15.47 & -4.21 & 3519 &  413 & 2.92 & 0.91 & 0.73 & 0.32 &  -63 \\ 
E5-2 003 & 14.0 &  49 & 13.76 & -2.86 & 3932 &  419 & 3.49 & 0.91 & 0.02 & 0.32 & -189 \\ 
E5-2 004 & 13.9 &  51 & 15.01 & -3.50 & 3808 &  414 & 3.29 & 0.91 & 0.39 & 0.32 &  -13 \\ 
E5-2 005 & 14.2 &  38 & 14.19 & -2.05 & 4328 &  415 & 3.85 & 0.91 & 0.40 & 0.32 &  -29 \\ 
E5-2 006 & 14.3 &  36 & 13.29 & -3.38 & 4071 &  418 & 3.87 & 0.91 & 0.45 & 0.32 &  157 \\ 
E5-2 009 & 14.9 &  38 & 14.49 & -2.26 & 4251 &  423 & 3.09 & 0.91 & 0.00 & 0.33 &    4 \\ 
E5-2 010 & 15.1 &  51 & 15.67 & -2.25 & 4434 &  416 & 3.91 & 0.91 & 0.18 & 0.32 &  -37 \\ 
E5-2 011 & 15.4 &  41 & 15.72 & -3.19 & 4613 &  415 & 3.94 & 0.91 & 0.16 & 0.32 &  -13 \\ 
E5-2 019 & 15.6 &  37 & 15.10 & -3.87 & 4131 &  415 & 3.69 & 0.91 & 0.11 & 0.32 &  -21 \\ 
E5-2 020 & 15.7 &  36 & 14.65 & -2.91 & 4066 &  420 & 3.79 & 0.91 & 0.42 & 0.32 &  -82 \\ 
E6-1 001 & 12.0 &  53 & 17.88 &  0.89 & 3687 &  413 & 3.17 & 0.91 & 0.60 & 0.32 &   84 \\ 
E6-1 002 & 12.8 &  54 & 17.95 & -1.80 & 3737 &  414 & 3.19 & 0.91 & 0.47 & 0.32 &  -59 \\ 
E6-1 003 & 13.3 &  48 & 18.92 & -1.56 & 3656 &  414 & 3.04 & 0.91 & 0.65 & 0.32 &    4 \\ 
E6-1 004 & 13.3 &  75 & 18.92 &  0.21 & 4155 &  413 & 3.68 & 0.91 & -0.12 & 0.32 &  -61 \\ 
E6-1 005 & 13.8 &  58 & 17.91 & -1.02 & 3974 &  414 & 3.47 & 0.91 & 0.33 & 0.32 & -218 \\ 
E6-1 006 & 13.8 &  44 & 19.45 & -1.53 & 3720 &  414 & 3.25 & 0.91 & 0.51 & 0.32 &  -38 \\ 
E6-1 008 & 14.8 &  38 & 18.40 & -1.13 & 3986 &  416 & 3.68 & 0.91 & 0.43 & 0.32 &   -4 \\ 
E6-1 009 & 14.9 &  55 & 17.56 & -0.47 & 4358 &  413 & 3.86 & 0.91 & 0.41 & 0.32 & -124 \\ 
E6-1 010 & 15.4 &  49 & 18.18 & -0.47 & 4590 &  417 & 3.63 & 0.91 & -0.04 & 0.32 & -105 \\ 
E6-2 001 & 13.0 &  66 & 19.06 & -2.98 & 3859 &  414 & 3.19 & 0.91 & 0.43 & 0.32 &   46 \\ 
E6-2 002 & 13.5 &  50 & 16.21 & -3.57 & 3759 &  413 & 3.26 & 0.91 & 0.39 & 0.32 &  -52 \\ 
E6-2 003 & 14.0 &  42 & 19.07 & -2.56 & 3687 &  414 & 3.07 & 0.91 & 0.63 & 0.32 &    1 \\ 
E6-2 004 & 14.0 &  42 & 16.54 & -3.67 & 3730 &  415 & 3.08 & 0.91 & 0.42 & 0.32 &  -10 \\ 
E6-2 005 & 14.8 &  36 & 17.49 & -2.98 & 3836 &  414 & 3.30 & 0.91 & 0.64 & 0.32 &   16 \\ 
E6-2 006 & 14.8 &  41 & 17.24 & -3.20 & 3898 &  419 & 3.47 & 0.91 & 0.53 & 0.32 &   71 \\ 
E6-2 008 & 15.2 &  45 & 18.12 & -2.94 & 4063 &  415 & 3.61 & 0.91 & 0.24 & 0.32 &  -71 \\ 
E6-2 009 & 15.1 &  42 & 17.80 & -4.55 & 4422 &  416 & 3.86 & 0.91 & 0.16 & 0.32 &  -49 \\ 
E7-1 001 & 10.8 &  62 & 20.46 & -2.29 & 3479 &  413 & 2.92 & 0.91 & 0.73 & 0.32 & -105 \\ 
E7-1 002 & 11.7 &  56 & 21.80 & -0.93 & 3662 &  413 & 2.96 & 0.91 & 0.56 & 0.32 &  -70 \\ 
E7-1 003 & 12.1 &  41 & 21.40 & -1.99 & 3591 &  414 & 2.90 & 0.91 & 0.76 & 0.32 &   35 \\ 
E7-1 004 & 12.3 &  45 & 20.28 & -1.08 & 3594 &  414 & 2.95 & 0.91 & 0.72 & 0.32 & -120 \\ 
E7-1 005 & 13.4 &  39 & 20.95 & -2.11 & 3888 &  417 & 3.41 & 0.91 & 0.39 & 0.32 &   23 \\ 
E7-1 006 & 13.5 &  52 & 22.73 & -0.77 & 3878 &  414 & 3.29 & 0.91 & 0.52 & 0.32 &  -29 \\ 
E7-1 007 & 13.3 &  64 & 20.67 & -2.39 & 3479 &  413 & 2.83 & 0.91 & 0.64 & 0.32 & -118 \\ 
E7-1 010 & 14.6 &  39 & 20.47 & -1.19 & 3663 &  416 & 2.78 & 0.91 & 0.41 & 0.32 & -135 \\ 
E7-1 022 & 14.8 &  38 & 20.98 & -0.17 & 4218 &  425 & 3.63 & 0.91 & 0.44 & 0.32 &   38 \\ 
E7-2 001 & 11.4 &  51 & 20.78 & -2.97 & 3524 &  413 & 2.85 & 0.91 & 0.74 & 0.32 & -189 \\ 
E7-2 002 & 14.3 &  50 & 20.20 & -4.42 & 3801 &  414 & 3.27 & 0.91 & 0.39 & 0.32 & -204 \\ 
E7-2 006 & 15.4 &  56 & 20.49 & -3.73 & 4313 &  416 & 3.76 & 0.91 & 0.17 & 0.32 &  -90 \\ 
N1-1 001 & 13.2 &  44 &  3.51 &  8.17 & 3644 &  414 & 2.99 & 0.91 & 0.68 & 0.32 &  -47 \\ 
N1-1 002 & 13.3 &  81 &  1.40 &  8.51 & 4198 &  413 & 3.85 & 0.91 & -0.91 & 0.32 &  -72 \\ 
N1-1 003 & 13.9 &  46 &  4.25 &  9.00 & 3844 &  416 & 3.36 & 0.91 & 0.48 & 0.32 & -128 \\ 
N1-1 004 & 14.1 &  37 &  3.89 &  8.04 & 3988 &  417 & 3.36 & 0.91 & 0.36 & 0.32 &   -3 \\ 
N1-1 005 & 13.6 &  45 &  3.50 &  9.63 & 3658 &  414 & 3.18 & 0.91 & 0.75 & 0.32 & -122 \\ 
N1-1 007 & 14.4 &  36 &  3.97 &  8.89 & 3851 &  415 & 3.22 & 0.91 & 0.61 & 0.32 &   79 \\ 
N1-1 045 & 12.1 &  84 &  3.64 &  7.22 & 4212 &  413 & 3.97 & 0.91 & -1.14 & 0.32 &  170 \\ 
N1-2 001 & 13.3 &  56 &  5.92 &  6.93 & 3929 &  415 & 3.52 & 0.91 & 0.40 & 0.32 &   57 \\ 
N1-2 002 & 13.3 &  51 &  6.67 &  7.99 & 3706 &  413 & 3.19 & 0.91 & 0.67 & 0.32 &   53 \\ 
N1-2 003 & 13.5 &  41 &  5.48 &  8.60 & 3689 &  413 & 3.21 & 0.91 & 0.65 & 0.32 &   60 \\ 
N1-2 004 & 13.9 &  47 &  6.33 &  7.68 & 3844 &  413 & 3.74 & 0.91 & 0.36 & 0.32 &  137 \\ 
N1-2 006 & 14.9 &  36 &  6.08 &  8.20 & 4204 &  420 & 3.86 & 0.91 & 0.39 & 0.32 &   30 \\ 
N1-2 016 & 16.7 &  35 &  4.86 &  7.50 & 4484 &  416 & 3.10 & 0.91 & 0.13 & 0.32 &  -83 \\ 
N2-1 001 & 12.1 &  49 &  5.79 & 11.57 & 3593 &  413 & 3.12 & 0.91 & 0.69 & 0.32 &   61 \\ 
N2-1 002 & 12.1 & 110 &  4.40 & 11.00 & 4358 &  413 & 3.92 & 0.91 & -1.06 & 0.32 &  223 \\ 
N2-1 003 & 12.8 &  43 &  3.59 & 11.53 & 4260 &  415 & 3.82 & 0.91 & -1.20 & 0.32 &   31 \\ 
N2-1 004 & 13.1 &  37 &  3.63 & 10.68 & 3909 &  415 & 3.41 & 0.91 & 0.18 & 0.32 &   48 \\ 
NE1-1 001 & 10.4 &  54 & 11.54 &  3.73 & 3558 &  414 & 3.00 & 0.91 & 0.80 & 0.32 &   55 \\ 
NE1-1 002 & 10.7 &  52 &  8.58 &  4.10 & 3447 &  414 & 2.78 & 0.91 & 0.90 & 0.32 & -141 \\ 
NE1-1 003 & 11.4 & 133 &  8.61 &  3.76 & 4125 &  413 & 3.98 & 0.91 & -1.27 & 0.32 & -119 \\ 
NE1-1 005 & 12.4 &  50 &  9.43 &  4.95 & 3517 &  413 & 2.93 & 0.91 & 0.89 & 0.32 & -135 \\ 
NE1-1 007 & 13.5 &  39 &  9.68 &  2.99 & 3710 &  414 & 3.21 & 0.91 & 0.65 & 0.32 &   59 \\ 
NE1-1 008 & 13.4 &  48 & 10.54 &  4.92 & 3625 &  415 & 3.02 & 0.91 & 0.81 & 0.32 & -106 \\ 
NE1-1 009 & 13.5 &  52 &  8.98 &  4.25 & 3683 &  413 & 3.27 & 0.91 & 0.56 & 0.32 &  -43 \\ 
NE1-1 010 & 13.9 &  43 & 10.80 &  3.67 & 3769 &  417 & 3.30 & 0.91 & 0.62 & 0.32 & -186 \\ 
NE1-1 011 & 14.2 &  46 &  9.36 &  3.15 & 3829 &  414 & 3.35 & 0.91 & 0.56 & 0.32 &   34 \\ 
NE1-1 012 & 14.5 &  41 & 11.11 &  4.41 & 4239 &  418 & 3.75 & 0.91 & 0.48 & 0.32 & -171 \\ 
NE1-1 013 & 14.8 &  41 & 11.08 &  4.79 & 3885 &  422 & 3.35 & 0.91 & 0.33 & 0.32 & -225 \\ 
NE1-1 014 & 15.3 &  38 & 11.51 &  2.86 & 4241 &  417 & 3.92 & 0.91 & 0.03 & 0.32 &   54 \\ 
NE1-1 018 & 15.1 &  39 & 10.37 &  3.57 & 4354 &  417 & 3.79 & 0.91 & 0.20 & 0.33 &  -41 \\ 
NE1-1 025 & 15.3 &  52 &  9.18 &  2.84 & 4224 &  419 & 3.66 & 0.91 & 0.55 & 0.32 &  -39 \\ 

\enddata
\tablenotetext{a}{Name, K magnitude, and RA \& DEC offset from Sgr A* from \citep{2015arXiv150407239S}.}
\tablenotetext{b}{Uncertainties include statistical uncertainties, interpolation uncertainties and systematic uncertainties added in quadrature.}
\label{tab:results}
\end{deluxetable*}

\end{document}